\documentclass[twocolumn]{aastex62}
\usepackage{amssymb}
\usepackage{graphicx}
\usepackage{color}
\usepackage{enumerate}

\newcommand{\SampleSize}{2,330} 

\newcommand\lamost{LAMOST}
\newcommand\apogee{APOGEE}
\newcommand\rave{RAVE}
\newcommand{\project}[1]{\emph{#1}}
\newcommand{\gaia}{\project{Gaia}}

\newcommand{\kepler}{\project{Kepler}}
\newcommand{\ktwo}{\project{K2}}
\newcommand{\logg}{\log_{10}[g\,({\rm cm\,s}^{-2})]}

\received{January X, 2019}
\revised{January X, 2019}
\accepted{January X, 2025}
\submitjournal{AAS Journals}

\shorttitle{Tidal interactions between binary stars drives lithium production in low-mass red giants}
\shortauthors{Casey et al.}

\begin{document}

\title{Tidal interactions between binary stars drives lithium production in low-mass red giants}

\correspondingauthor{Andrew R. Casey}
\email{andrew.casey@monash.edu}

\author[0000-0003-0174-0564]{Andrew~R.~Casey}
\affil{School of Physics \& Astronomy, 
       Monash University, Clayton 3800, Victoria, Australia}
\affil{Faculty of Information Technology,
	   Monash University, Clayton 3800, Victoria, Australia}
	   
\author[0000-0002-9017-3567]{Anna~Y.~Q.~Ho}
\affil{Cahill Center for Astrophysics, California Institute of Technology, 
       MC 249-17, 1200 E California Blvd, Pasadena, CA, 91125, USA}
\affil{Max-Planck-Institut f\"ur Astronomie, K\"onigstuhl 17, D-69117 
       Heidelberg, Germany}       

\author{Melissa~Ness}
\affil{Department of Astronomy, Columbia University, 550 W 120th St, 
	   New York, NY 10027, USA}
\affil{Max-Planck-Institut f\"ur Astronomie, K\"onigstuhl 17, D-69117 
       Heidelberg, Germany}
       
\author{Hans-Walter~Rix}
\affil{Max-Planck-Institut f\"ur Astronomie, K\"onigstuhl 17, D-69117 
       Heidelberg, Germany}
              
\author{George~C.~Angelou}
\affil{School of Physics \& Astronomy, 
       Monash University, Clayton 3800, Victoria, Australia}
\affil{Max Planck Institut f\"ur Sonnensystemforschung, Justus-von-Liebig-Weg 3, 37077 
       G\"ottingen, Germany}
\affil{Max-Planck Institut f\"ur Astrophysik, Karl-Schwarzschild-Str. 1, D-85741 
       Garching, Germany}
	  
\author{Saskia~Hekker}
\affil{Max Planck Institute for Solar System Research,
	   SAGE research group,
	   Justus-von-Liebig-Weg 3,
	   37077 G\"ottingen, Germany}
\affil{Stellar Astrophysics Centre, Dept. of Physics and Astronomy, 
	   Aarhus University, Ny Munkegade 120, 8000 Aarhus C, Denmark}

\author{Christopher~A.~Tout}
\affil{School of Physics \& Astronomy, 
       Monash University, Clayton 3800, Victoria, Australia}
\affil{Institute of Astronomy, University of Cambridge, Madingley Road, 
       Cambridge CB3 0HA, UK}

\author{John~C.~Lattanzio}
\affil{School of Physics \& Astronomy, 
       Monash University, Clayton 3800, Victoria, Australia}
       
\author{Amanda~I. Karakas}
\affil{School of Physics \& Astronomy, 
       Monash University, Clayton 3800, Victoria, Australia}
       
\author{Tyrone~E.~Woods}
\affil{Institute for Gravitational Wave Astronomy and School of Physics and Astronomy, 
	   University of Birmingham, Birmingham, B15 2TT, UK}
\affil{School of Physics \& Astronomy, 
       Monash University, Clayton 3800, Victoria, Australia}
       
\author[0000-0003-0872-7098]{Adrian~M.~Price-Whelan}
\affil{Department of Astrophysical Sciences, Princeton University, 
       4 Ivy Lane, Princeton, NJ 08544, USA}

\author[0000-0001-5761-6779]{Kevin~C.~Schlaufman}
\affil{Department of Physics and Astronomy, Johns Hopkins University, 
	   3400 N Charles St., Baltimore, MD 21218, USA}

\begin{abstract}
Theoretical models of stellar evolution predict that most of the lithium inside
a star is destroyed as the star becomes a red giant. 
However, observations reveal that about 1\% of red giants are peculiarly rich
in lithium, often exceeding the amount in the interstellar 
medium or predicted from the Big Bang. With only about 150 lithium-rich
giants discovered in the past four decades, and no distinguishing properties 
other than lithium enhancement, the origin of lithium-rich 
giant stars is one of the oldest problems in stellar astrophysics. Here we 
report the discovery of \SampleSize\ low-mass ($1\,\,{\rm to}\,\,3\,M_\odot$) lithium-rich 
giant stars, which we argue are consistent with internal lithium production 
that is driven by tidal spin-up by a binary companion. Our sample reveals
that most lithium-rich giants have helium-burning cores ($80^{+7}_{-6}\%$), and that the frequency
of lithium-rich giants rises with increasing stellar metallicity. We find that 
while planet accretion may explain some lithium-rich giants, it 
cannot account for the majority that have helium-burning cores. We rule out most other 
proposed explanations as the primary mechanism for lithium-rich giants, including all
stages related to single star evolution.
Our analysis shows that
giants remain lithium-rich for only about two million years. A prediction from
this lithium depletion timescale is that most lithium-rich giants with a helium-burning
core have a binary companion.
\end{abstract}


\vspace{2.5em}
\section{Introduction} \label{sec:intro}
Stellar evolution theory suggests that material from inner layers,
where the element composition has been altered by nuclear reactions, is
dredged up to the surface when a star evolves to become a red giant.
The surface abundances of certain elements are predicted to change as a
consequence of this process. These elements include helium, carbon, nitrogen, and 
an approximate 95\% drop in lithium content \citep{Iben_1967}. Observations have 
repeatedly confirmed these predictions \citep{Lambert_1981,Gilroy_1989,
Kirby_2016}, yet also revealed rare examples of otherwise normal giant stars
with high surface lithium abundances \citep[e.g.,][]{Martell_2013}. In some giants the lithium 
content is higher than what is inferred for the surrounding interstellar medium, indicating that
lithium cannot just be preserved: there must be an accretion or production 
mechanism \citep{Charbonnel_2000}. However, the temperature required to 
produce lithium is also sufficient to destroy it: helium isotopes must be 
fused together at high temperatures to produce beryllium-7, and beryllium-7
must be transported to cooler regions where lithium can form by 
electron capture without being immediately destroyed by proton capture \citep{Cameron_1971}. 
These strict requirements make lithium extremely sensitive to the structure
and mixing inside a star. Standard theoretical models cannot produce
appreciable net amounts of lithium for red giant branch stars. This has prompted several descriptions of
non-standard mixing \citep{Sweigart_1979,Lattanzio_2014,Fekel_1993,Charbonnel_1995,
Sackmann_1999,Charbonnel_2000,Denissenkov_2003}, as well as hypotheses that
lithium production is associated with a specific stage of stellar
evolution \citep{Charbonnel_2000,Kumar_2011,Lattanzio_2014}, or the result 
of external phenomena \citep{Siess_1999,Andrievsky_1999,Denissenkov_2004}. 
The lack of evolutionary phase information for a large sample of
lithium-rich giants has until now prohibited any empirical constraints on why, where, 
and when lithium production occurs, and for how long stars remain lithium-rich.

\section{Methods and Results} \label{sec:methods}

\subsection{Spectroscopy}

\begin{figure*}
    \includegraphics[width=\textwidth]{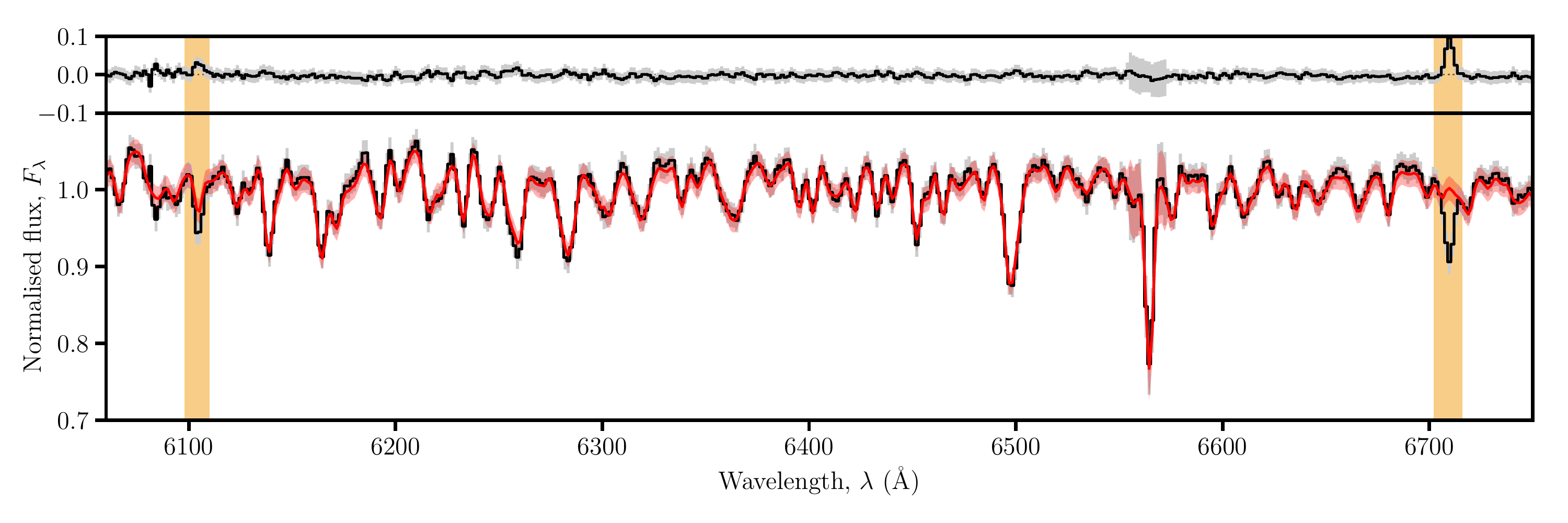}
        \caption{A portion of the \lamost\ spectrum and best-fitting model for an example lithium-rich giant star, J055640.1+144534.
        The data are shown in black and $1\sigma$ flux uncertainties are shaded in grey. A data-driven model of a lithium-normal star is shown in red, where the quadrature sum of model and data uncertainties are shaded in red. We mark the regions surrounding the 6104\,\AA\ and 6707\,\AA\ lithium transitions where we searched for significant residuals.}
    \label{fig:spectrum}
\end{figure*}

We identified candidate lithium-rich giant stars using public spectra from 
the \lamost\ survey \citep[Data Release 2;][]{Luo_2015}. Specifically, we searched for significant
deviations between the continuum-normalised rest-frame spectrum 
and a best-fit spectrum from a data-driven model \citep[Figure~\ref{fig:spectrum};][]{Ho_2017a,Ho_2017b}. We applied a gaussian matched filter to the 
residuals between the model and the data at the 6707\,\AA\ lithium doublet 
and the 6104\,\AA\ subordinate line.  We identified 4,558 candidate 
lithium-rich giants by requiring a $3\sigma$ deviation in either region. 
We visually inspected the spectrum and best-fitting model for every 
lithium-rich giant star candidate, twice. We discarded any candidate that 
showed evidence of being a false positive, including spectra with very low
signal-to-noise ratios or data reduction issues, as well as any candidate where the lithium 
deviations were narrower than the expected spectral resolution.

The evolutionary track for low-mass pre-main-sequence stars overlaps with
the sub-giant phase in stellar effective temperature and surface 
gravity \citep{Dotter_2016,Choi_2016}. Consequently we discarded 302 
lithium-rich sub-giant candidates because they showed evidence of being 
young stars, either through chromospheric activity indicated by emission 
in H$\alpha$, or significant photometric variability indicating star 
spots \citep{McQuillan_2014}. We found that most stars with these signatures were
spatially concentrated in known young star-forming regions or at low 
absolute Galactic latitudes. 

%In Extended Data Figure 2 we show the spatial
%coverage of \lamost, and the sky positions of all lithium-rich giant stars
%in our sample.

\begin{figure*}[t]
    \includegraphics[width=\textwidth]{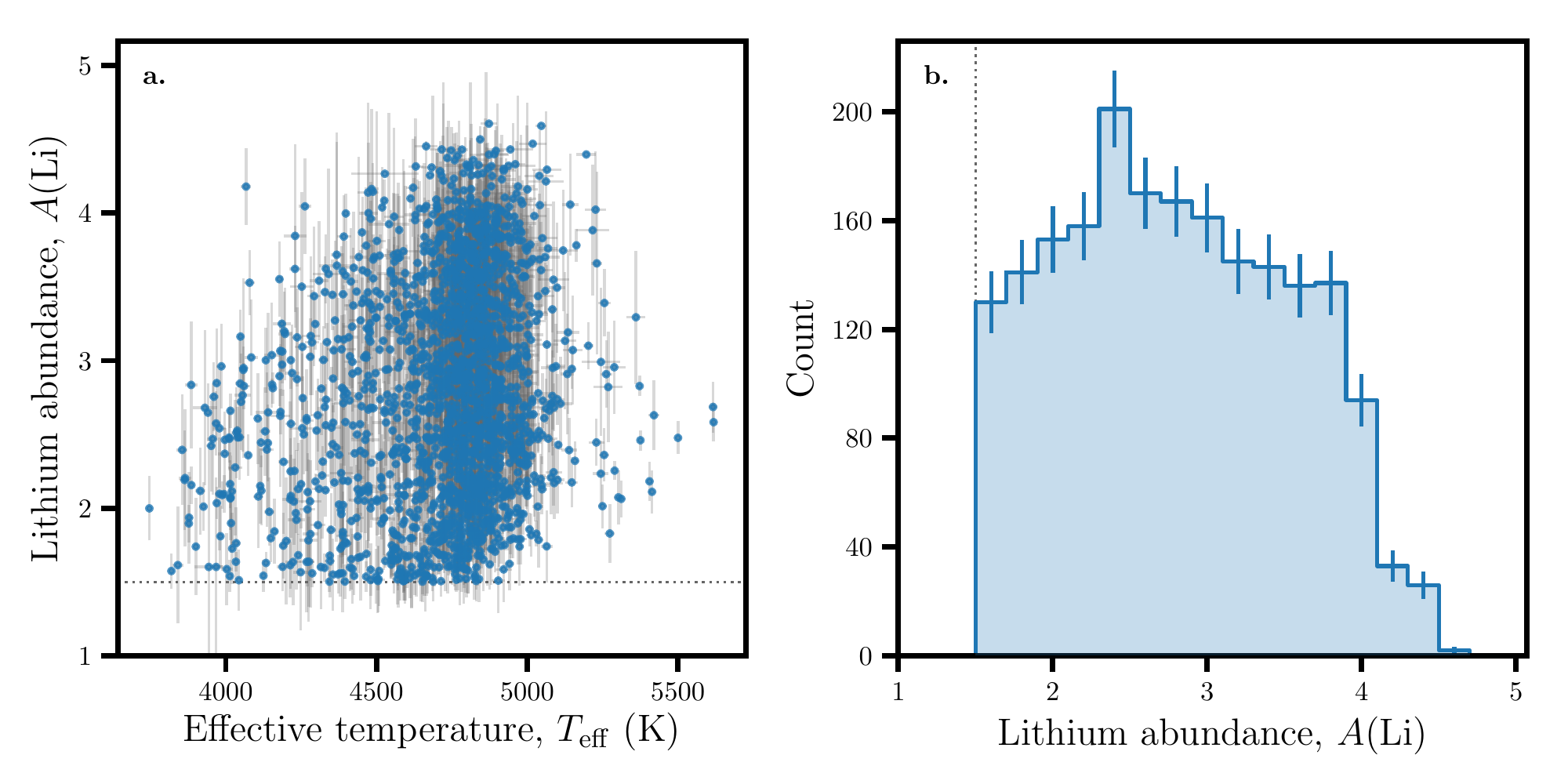}
	\caption{Distribution of measured lithium abundances. \textbf{a.} Surface lithium abundances against stellar effective temperature for all \SampleSize\ lithium-rich giant stars discovered in \lamost. It is plausible that not all lithium-rich giants with $T_{\rm eff} > 5000\,{\rm K}$ and $A({\rm Li}) = 1.5$ to $2.5$ are identified by our matched filter. \textbf{b.} The distribution of measured lithium abundances.  The dotted line in both panels represents the defining limit of $A({\rm Li}) > 1.5$ for a lithium-rich giant star.}
    \label{fig:li_distribution}
\end{figure*}

Using the stellar parameters ($T_{\rm eff}$, $\log_{10}{g}$, $[{\rm Fe/H}]$)
derived from \lamost\ spectra \citep{Ho_2017a}, we synthesised the 6707~\AA\ lithium 
doublet and surrounding region and determined the best-fitting lithium abundance for 
each star. We used MARCS spherical model photospheres \citep{Gustafsson_2008}, the
VALD database of transitions \citep{Piskunov_1995}, and the \texttt{iSpec} 
\citep{Blanco_Cuaresma_2014} wrapper of the \texttt{SME} synthesis package \citep{Valenti_1996}.
Using the standard nomenclature of 
$A({\rm Li}) = \log_{10}(N_{\rm Li}/N_{H}) + 12$, where $N_X$ refers to the
number density of atoms of a species, we excluded 15 stars with $A({\rm Li}) < 1.5$
as being lithium-normal. Our distilled sample contains \SampleSize\ 
lithium-rich giant stars. We show the distribution of $A({\rm Li})$ for these 
lithium-rich giants in Figure~\ref{fig:li_distribution}.

We find that two of our lithium-rich giants are rediscoveries:
SDSS~J0652+4052 and SDSS~J0654+4200 \citep{Martell_2013}. The stellar parameters and lithium 
abundances ($T_{\rm eff}$, $\log_{10}g$, $[{\rm Fe/H}]$, $A({\rm Li})$) we 
derive are all consistent within the joint $2\sigma$ uncertainty between this
work and the literature, with most measurements agreeing within $1\sigma$ of 
the quoted uncertainty in either study. In particular we find 
$A({\rm Li}) = 3.47 \pm 0.19$ for SDSS~J0654+4200, in good agreement 
with the previously reported value of $A({\rm Li})~=~3.3~\pm~0.2$,
and $A({\rm Li})~=~3.26~\pm~0.08$ for SDSS~J0652+4052, 0.04 below 
the literature value. We also note that \lamost\ obtained a high signal-to-noise
ratio spectrum for another known lithium-rich giant star \citep[SDSS~J0304+3823;][]{Martell_2013}, 
but this was not included in our sample because the residuals surrounding the
lithium doublet at 6707\,\AA\ reached only $2.7\sigma$, and did not meet our $3\sigma$
threshold for detection.

The data-driven model we employed also provides estimates of [C/H] and [N/H] abundance
ratios for all \lamost\ spectra. We found that 30 of our lithium-rich 
($A({\rm Li}) > 1.5$) giants at the base of the giant branch 
($\logg > 3.2$) have $[{\rm C/N}] > 0$, which indicates that first dredge-up may not have finished and therefore lithium is not expected to be fully depleted.
We include these candidates in our sample but caution that first 
dredge-up may not have finished and this may explain their high
lithium content. We estimated stellar masses from stellar parameters and 
[C,N/H] abundance ratios for 1,374 of our lithium-rich giants, where their 
stellar parameters and abundance ratios are in the valid range for existing 
empirical relationships \citep{Martig_2016}. These inferred masses indicate 
that most of our sample are low-mass (1 to 3$M_\odot$) red giant stars.

\subsection{Asteroseismology}

\begin{figure*}
	\includegraphics[width=\textwidth]{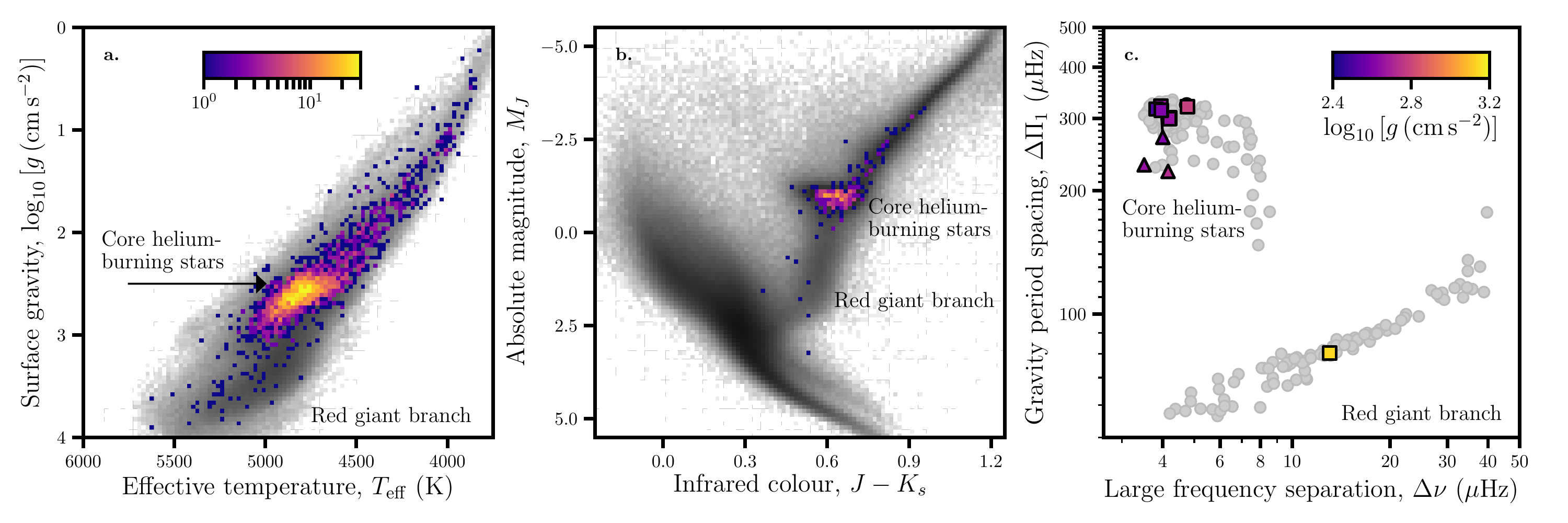}
	\caption{
		{Most lithium-rich giants have helium-burning cores.}
		\textbf{a.} Stellar parameters for all giants in \lamost\ (shown as logarithmic density in grey) and \SampleSize\ lithium-rich giants coloured by logarithmic density.
		\textbf{b.} Infrared colour and absolute magnitude for 240 lithium-rich giants
		with \gaia\ parallaxes. The density colour scale in panel (b) is matched to panel (a).
		The entire \gaia--TGAS sample is shown in grey \citep{Gaia-Collaboration_2018,Anderson_2018}.
		\textbf{c.} Asteroseismic period spacings for lithium-rich giants in the
		\kepler\ field.  Marker shapes (circles \citep{Mosser_2012}, squares \citep{Stello_2013} and triangles \citep{Vrard_2016}) indicate the literature source. For context, we show 
		asteroseismic properties of typical giant stars in grey \citep{Mosser_2012}.}
	\label{fig:figure1}
\end{figure*}

Asteroseismology confirms the results we derive from spectroscopy. 
The evolutionary states for 23 of our lithium-rich giants could be unambiguously determined
using high-quality light 
curves from the \kepler\ space telescope \citep[Figure~\ref{fig:figure1}c;][]{Mosser_2012,Stello_2013,Vrard_2016}, 
which confirms they are low-mass red giant branch stars, and where at 
least 21 are found to be core-helium burning stars (perhaps 22; the classification of one star is disputed).
Another 2 lithium-rich giants have useful light curves obtained during the \kepler/\ktwo\ mission. 
Our asteroseismic analysis of those \kepler/\ktwo\ light curves reveals that these two upper giant 
branch stars are first ascent giants, with likelihood ratios of about 100 when compared to core
helium-burning or asymptotic giant branch phases \citep{Hekker_2017}. Ignoring selection effects associated with the \kepler\ and \ktwo\ missions, this asteroseismic sample of 25 suggests that the fraction of lithium-rich giants with helium burning cores is about $f_{\rm CHeB} = 0.84$ to $0.88$ ($21/25$ to $22/25$).

We cross-matched the complete \lamost\ catalog with a literature source of asteroseismic
properties \citep[$\Delta\nu$, $\Delta\Pi_1$;][]{Vrard_2016}, which revealed 1,365 stars that 
have both high quality \lamost\ spectra and high fidelity asteroseismic labels. With these 
data as a training set, we modelled the flux in each \lamost\ pixel as a second-order quadratic function
of the stellar labels ($T_{\rm eff}$, $\logg$, $[{\rm Fe}/{\rm H}]$, $\delta\nu$, and $\Delta\Pi_1$)
and a single noise term per pixel. At the test step we classified a star as having a helium-burning core if 
the estimated $\Delta\Pi_1 > 150\,{\rm s}$. We performed cross-validation experiments ($N_\textrm{trials} = 10$) 
where we used a random 80\% of the training set as the labelled set and the remaining 20\% formed as a validation 
set. From these experiments we find that our model can identify core-helium burning stars directly
from \lamost\ spectra with an accuracy (recall) of 93.4\% (precision 96.9\%; F-measure 0.95). 

We applied this classifier to all \SampleSize\ lithium-rich giants in \lamost\ and find that the
fraction of lithium-rich giants with helium-burning cores is $f_{\rm CHeB} = 0.80^{+0.07}_{-0.06}$ 
(95\% confidence interval). This result is only negligibly dependent on the \kepler\ selection function, and is fully consistent with what we find from the small sample of lithium-rich giants 
with reported asteroseismic properties ($f_{\rm CHeB} = 0.84$ to $0.88$). For these reasons, 
we take the fraction of lithium-rich giants with helium-burning cores to be 
$f_{\rm CHeB} = 0.80^{+0.07}_{-0.06}$ for the remainder of this work.

\vspace{2em}
\subsection{Timescales}

The relative timescales of different stages of stellar evolution are known with
high precision, and are not model dependent. 
Without yet prescribing a mechanism for lithium 
enrichment, our sample size allows us to infer when stars become lithium-rich,
and how long they remain lithium-rich. We modelled the expected distribution 
in stellar parameters using evolutionary tracks \citep{Dotter_2016,Choi_2016}
for three different scenarios where giants could become lithium-rich:
\begin{enumerate}[(a)]
	\item at the luminosity bump \citep[e.g.,][]{Charbonnel_2000},
	\item at the tip of the red giant branch \citep[e.g.,][]{Lattanzio_2014}, or
	\item either during the core-helium burning phase or at a random time on the giant branch.
\end{enumerate}

For each lithium-rich giant star we selected the closest evolutionary track in 
mass and metallicity \citep{Dotter_2016,Choi_2016}. Lithium-rich giants without 
estimated masses were excluded from this analysis, and we used asteroseismic masses 
where available in preference to masses inferred from carbon and nitrogen abundances. 
For each tested scenario we assign a point along the selected evolutionary track 
where the giant would become lithium-rich, and a time for which it remained 
lithium-rich. Hereafter we refer to this timescale as the lithium depletion timescale. 
By combining the lithium-rich sections of each track, we calculate the normalised distribution 
of stellar parameters ($T_{\rm eff}$, $\log_{10}{g}$) we would expect to 
observe for lithium-rich giant stars. For example, if stars become lithium-rich 
at the luminosity bump and only remain lithium-rich for an instant then the expected
distribution in $\log_{10}{g}$ would only show stars close to the luminosity bump.
But if stars remain lithium-rich for say $10^8\,{\rm yr}$, long enough to evolve 
beyond the tip of the giant branch, then we would expect lithium-rich giants 
throughout the upper red giant branch, and some fraction of them to have
helium-burning cores.
Lithium depletion timescales between $10^4\,{\rm yr}$ and $10^8\,{\rm yr}$ were 
considered for all scenarios, and in each case we convolved the expected distribution in stellar 
parameters with the median observational uncertainties.

\begin{figure*}[t!]
	\includegraphics[width=\textwidth]{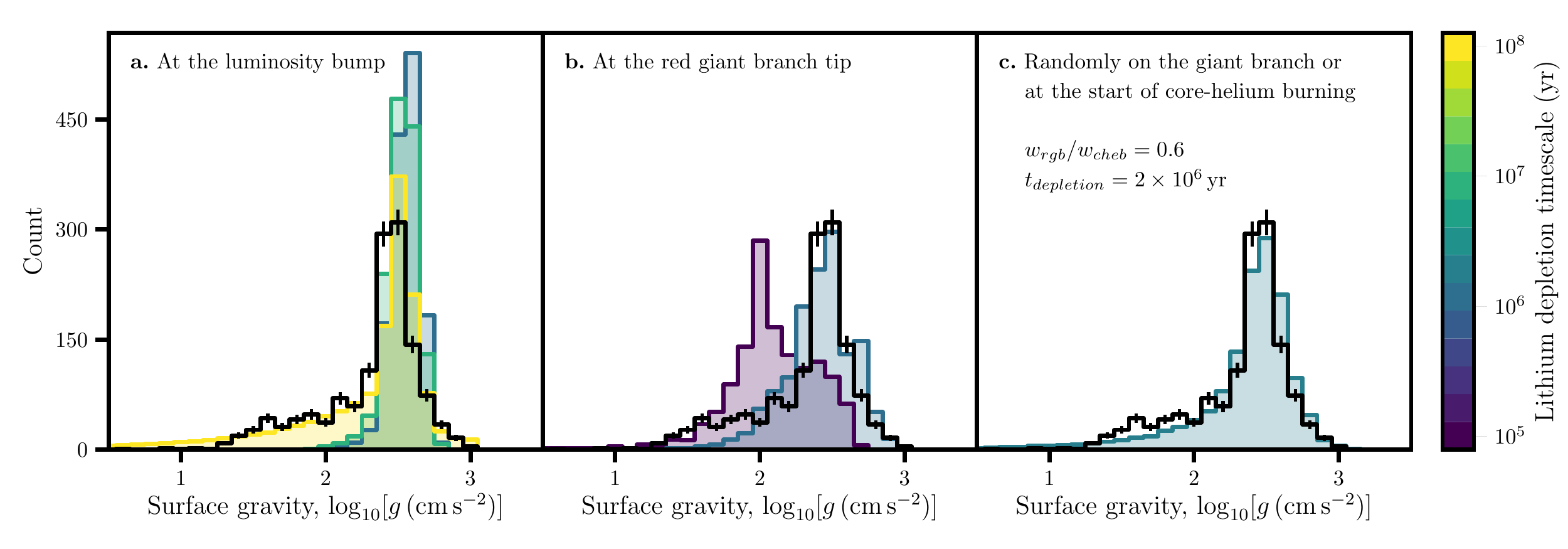}
	\caption{{The data are consistent with stars becoming lithium-rich at the start of core helium-burning, or at a random time on the red giant branch.} The observed distribution in $\logg$ is shown in black (all panels) for 1,374 lithium-rich giants with estimated masses, and error bars indicate the relative standard deviation of the number of stars per bin. The expected distribution for lithium production scenarios is shown in each panel. Scenario (a) and (b) cannot produce enough core-helium burning lithium-rich giants for any depletion timescale: in Scenario (a) at most 40\% of lithium-rich giants can have helium-burning cores, half the rate we observe $80^{+7}_{-6}$\,\%. In panel (c) we show the best fitting lithium depletion timescale from a grid search (see Section~\ref{sec:methods}).}
	\label{fig:evtracks}
\end{figure*}

In scenario (a), lithium 
production takes place at the luminosity bump on the giant branch. We identify the bump 
as the first luminosity reversal (brightness decrease) that occurs on the giant branch in the evolutionary track.
We find that lithium depletion timescales of at least $10^8\,{\rm yr}$ are 
required to produce lithium-rich giants with helium-burning cores in this scenario (Figure~\ref{fig:evtracks}a), simply because this is the typical timescale needed for
lithium-rich giants to evolve from the luminosity bump to the core helium-burning phase.
However, any timescale of $10^8\,{\rm yr}$ (or longer) results in at most only 40\% of lithium-rich giants having helium-burning cores, half the $80^{+7}_{-6}$\,\% we infer (95\% confidence interval).
This is because the time to evolve from the luminosity bump to the tip of the red giant branch is at least as long as the lifetime of core-helium burning.

In scenario (b) giants become lithium-rich at the tip of the red giant branch. At this point 
the star contracts rapidly, shrinking by a factor of ten in radius in less than $10^4\,{\rm yr}$,
but taking of order $10^6\,{\rm yr}$ until the star is fully established in the core-helium burning stage (i.e., a so-called red clump star). For this reason, lithium depletion timescales of at least $10^6\,{\rm yr}$ are necessary to account for most lithium-rich giants being stable core-helium burning stars. With this timescale (or longer), there is a paucity of lithium-rich stars descending from the tip of the giant branch. In fact, \emph{with any timescale}, scenario (b) cannot explain the $20^{+6}_{-7}$\,\% of lithium-rich giants that we find to be first ascent red giant branch stars, or similar examples known in the literature \citep[e.g.,][]{Kirby_2016}. If scenario (b) is the predominant mechanism for lithium-rich giants, then another pathway is required to explain first ascent red giant branch stars that are lithium-rich.

\begin{figure}
	\includegraphics[width=0.5\textwidth]{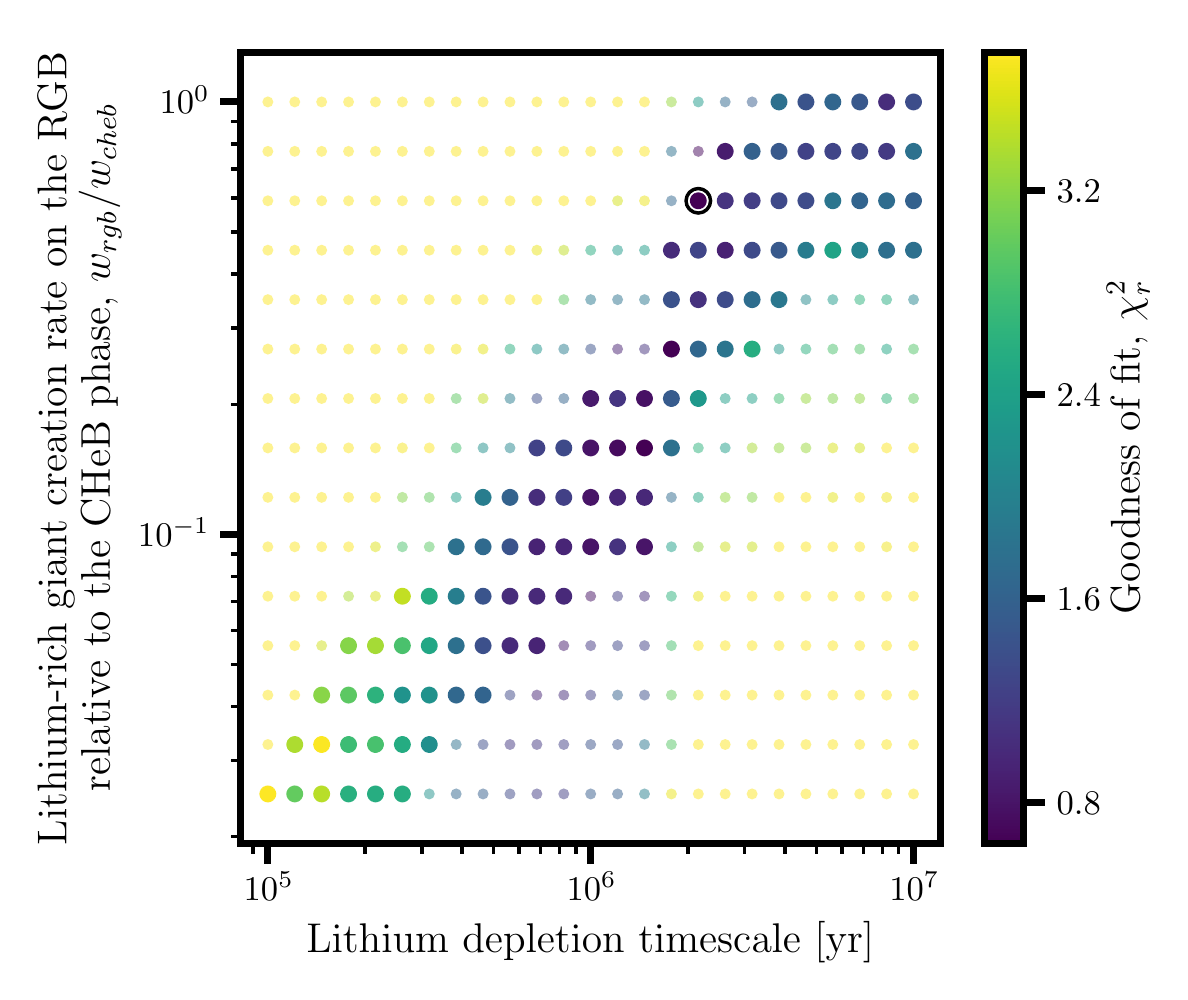}
	\caption{The preferred  lithium depletion timescale in scenario (c) is about $10^6$\,yr. Goodness of fit (reduced $\chi^2$) for a grid search of lithium depletion timescales and weighting ratios. Semi-transparent points indicate that the combination of weight and depletion time predicted core-helium burning fractions that are outside our 95\% confidence interval of $80^{+7}_{-6}$\,\%, and therefore inconsistent.  The preferred model ($w_\mathrm{rgb}/w_\mathrm{cheb} = 0.6$, $t_{depletion} = 10^{6.3}\,{\rm yr}$ or $2\times10^6\,{\rm yr}$), with $\chi_r^2 = 0.6$, is marked.}
	\label{fig:grid_search}
\end{figure}

Scenarios (a) and (b) represent the significant stellar evolution events that occur on the red giant branch. However, no timescale in either scenario provides an adequate explanation for the data. This would indicate that there is not a single phase of stellar evolution where significant internal lithium production occurs. For these reasons we considered a third scenario (c) where stars become lithium-rich at a uniformly random time from just before the luminosity bump until the tip of the giant branch, or they become
lithium-rich when they reach the stable core helium-burning phase. This scenario required us to introduce a relative weighting between the rates of creation of lithium-rich giants at the start of core-helium burning, relative to those created at some point on the red giant branch. If the lithium depletion timescale were short (i.e., much less than $10^6\,{\rm yr}$) then this weighting would exactly reproduce the observed fraction of core-helium burning stars relative to red giant branch stars. However, when the lithium depletion timescale is longer than the time a star takes to evolve from the red giant branch to the red clump, then some red clump giants we observe may have become lithium-rich on the red giant branch and simply remained lithium-rich until we observe them as red clump stars. To account for the random onset time of lithium production  for stars on the giant branch, we made 1,000 Monte Carlo draws from a uniform distribution in time for each observed star, where the lower bound is the time of the luminosity bump and the upper bound is the time of the red giant branch tip. In Figure~\ref{fig:grid_search} we show the goodness-of-fit $\chi_r^2$ from a grid search of trialled weights and depletion timescales for scenario (c). 

If we only consider lithium depletion timescales and weighting fractions that are consistent with our 95\% confidence interval of fraction of core helium burning stars ($f_\mathrm{CHeB} = 0.80^{+0.07}_{-0.06}$), we find that a lithium depletion timescale of about $2\times 10^{6}\,{\rm yr}$ with a relative formation rate of red giant branch stars to core-helium burning stars of 0.6:1.0 ($w_{RGB}/w_{CHeB} = 0.6$) is preferred with $\chi_r^2 = 0.6$, and provides reasonable agreement with the data (Figure~\ref{fig:evtracks}c). For comparison, $\chi_r^2$ values between 1.9 and 8.4 were found for all timescales considered in scenario (a), and between $\chi_r^2$ = 1.5 and 10 for those in scenario (b).
In Section~\ref{sec:discussion} we discuss physical mechanisms that are consistent with scenario (c).

\section{Discussion} \label{sec:discussion}

The main astrophysical parameters for the \SampleSize\ lithium-rich giants
that we discovered are shown in Figure \ref{fig:figure1}.  Our sample size
is some $100$ times bigger than the largest study to date \citep{Martell_2013}.
The stellar parameters we derive from spectroscopy suggest that $80^{+7}_{-6}$\% of lithium-rich 
giants have helium-burning cores, an analysis that is confirmed through independent expert
asteroseismic analyses.

\begin{figure}
	\centering
	\includegraphics[width=0.5\textwidth]{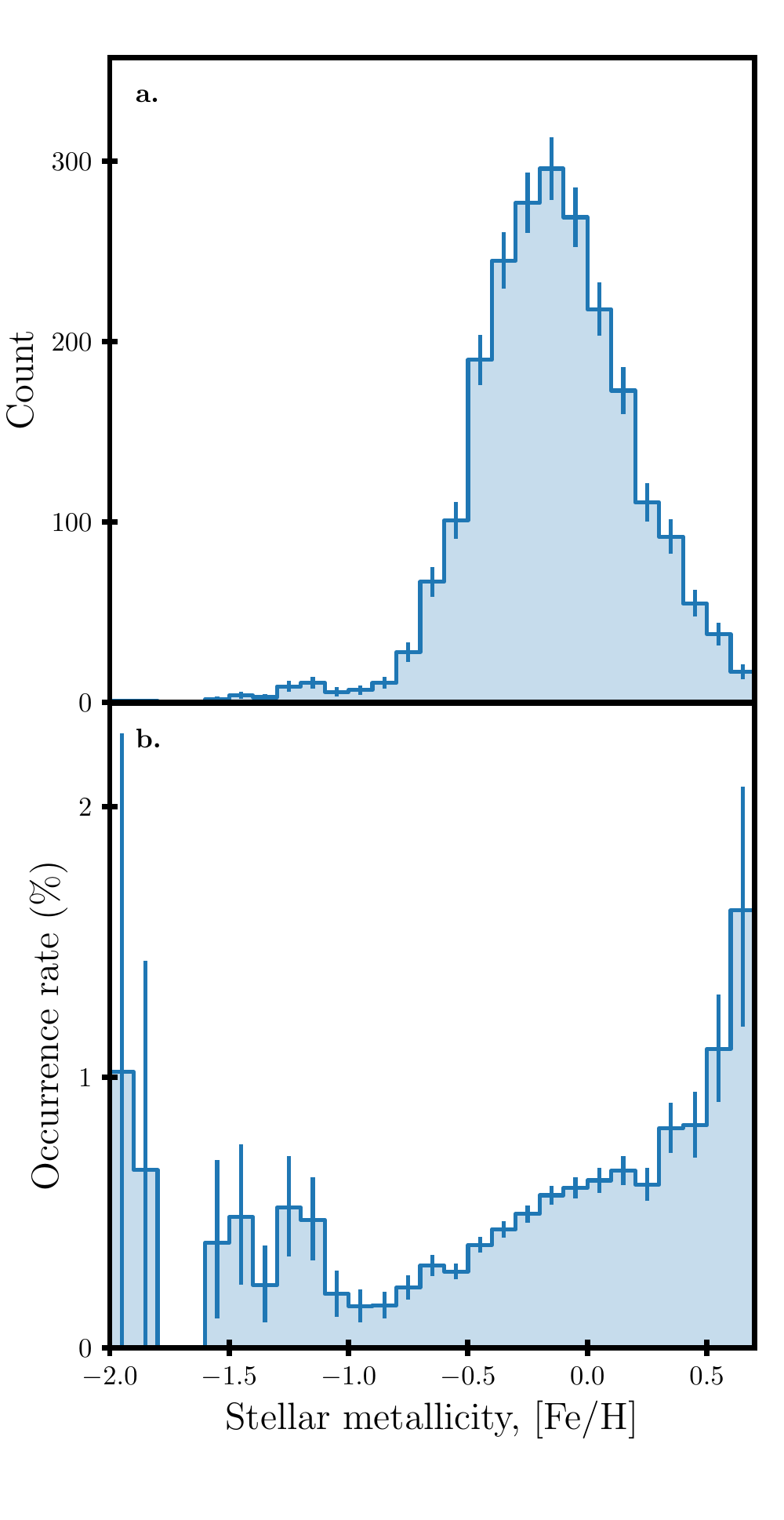}
	\caption{
		Lithium-rich giants are more frequent at higher metallicities.
		\textbf{a.} Metallicity distribution for all
		\SampleSize\ lithium-rich giant stars. Error bars represent the standard deviation of counts per bin.
		\textbf{b.} The occurrence 
		rate of lithium-rich giants with stellar metallicity. Error bars represent standard deviation of lithium-rich giants relative to number of stars per bin.}
	\label{fig:mdf_frequency}
\end{figure}

We find that lithium-rich giant stars occur more frequently with higher stellar 
metallicity ($[{\rm Fe/H}]$; Figure~\ref{fig:mdf_frequency}). This result 
reconciles tension between the low frequencies of lithium-rich giants reported 
in metal-poor environments with well-understood completeness statistics \citep{Kirby_2016} 
(e.g., $0.3 \pm 0.1$\% for isolated systems with $[{\rm Fe/H}] \lesssim -0.8$), 
as compared to field studies of metal-rich stars \citep{Brown_1989} (e.g., $1$ to $2$\%). 
This result has not been observed elsewhere likely because of the heterogeneous and 
serendipitous nature of lithium-rich giant star discoveries. Historically the 
discovery of a single lithium-rich giant star has warranted peer-reviewed 
publication\footnote{The literature compilation of lithium-rich giants by \citet{Casey_2016} shows that 73\% of publications announcing the discovery of lithium-rich giant(s) reported only 1 or 2 new lithium-rich giants.}, making it non-trivial to separate any metallicity 
relationship (or any other observable) from compounding selection effects.

\subsection{Selection effects}

We critically evaluated whether the \lamost\ target selection function could 
make us more or less likely to observe a lithium-rich giant star, and whether
the selection function could contribute to the increasing frequency of lithium-rich
giants we find with higher stellar metallicity. The \lamost\ target selection function
is a conglomerate of target selections for many simultaneous surveys. An
optimisation strategy is employed to maximise the number of fibres allocated
to potential targets (from any survey) for a single tiling plate, given physical
constraints such as fibre collisions. Given that the true distribution (or 
frequency) of lithium-rich giant stars is not known, these two facts conspire 
to prohibit us from directly inverting the \lamost\ selection function to 
understand biases. 

Some qualitative statements can be made despite this limitation. There
is no other observable property common to all lithium-rich giant stars that makes
them clearly distinguishable from lithium-normal stars. While some
lithium-rich giants do show excesses in infrared magnitudes, the LAMOST/LEGUE 
target selection function \citep{Carlin_2012} only makes use of the optical $g-r$
colour and an $r$ magnitude colour cut that is extended to match target 
densities for particular locations on the sky. Therefore, there is nothing
obvious in the \lamost\ selection function that could conceivably bias us towards,
or against, selecting lithium-rich giant stars. We also note that comparisons of 
the \lamost\ red giant branch sample with mock catalogues of the Milky Way 
do not show a significant bias in population properties \citep{Liu_2017}.

If we discard selection effects within \lamost\ as negligible, then there may still
be lithium-rich giant stars that we do not detect due to a weakened Li 
line at higher effective temperatures. This is shown in Figure~\ref{fig:li_distribution},
which indicates a possible lack of lithium-rich giants with $A({\rm Li}) = 1.5$ to $2.0$
and about $T_{\rm eff} > 5000\,{\rm K}$. It is possible that our $3\sigma$ detection
threshold means that we do not discover all lithium-rich sub-giant stars with $A({\rm Li}) = 1.5$ to $2.0$, which could imply that we do not detect all sub-giants that have become lithium-rich due to the engulfment of a close-in giant planet \citep{Casey_2016}. However, because our timescale modelling begins near the luminosity bump, if a selection effect is present it will not
affect our inferences on the lithium depletion timescale.

\subsection{Mechanisms and Interpretation}

Given a lithium depletion timescale and an occurrence rate of lithium-rich giants,
we can estimate the typical rate at which lithium-rich giants form 
($\dot N_{\rm{formation}} = N_{\rm{objects}}/\Delta t_{\rm{lifetime}}$). 
We assume a constant star formation rate of $2\,M_\odot\,{\rm yr}^{-1}$ and
consider stellar masses between $0.1\,M_\odot$ and $100\,M_\odot$ when weighting
a typical initial mass function \citep{Kroupa_2001} to find a birth rate of giant stars in the Milky Way (the main-sequence turn-off rate).
For an evolutionary track \citep{Dotter_2016,Choi_2016} of a $1.3\,M_\odot$ 
solar-metallicity star, the mean of our sample, the lifetime between $\logg = 3.2$
and the end of the core-helium burning phase is about $250\,{\rm Myr}$. 
Assuming a steady-state system, this implies that the number of giant stars in the Milky Way with 
$\logg < 3.2$ is about $7.5\times10^7$. Taking a mean fraction of 0.7\% 
lithium-rich giant stars, and the 305,793 giant stars in \lamost\ with 
$\logg < 3.2$, we estimate that there are about 527,100 lithium-rich 
giant stars in the Milky Way.  Taking $2\times10^{6}\,{\rm yr}$ as the 
lithium depletion timescale, this provides us with a formation rate of 
$\dot N_{\rm formation} = 0.3\,{\rm yr}^{-1}$.

This rate excludes merged binary stars \citep[$0.01\,{\rm yr}^{-1}$;][]{Andrievsky_1999}
and the engulfment of brown dwarfs \citep{Siess_1999} as the principal explanation
for lithium-rich giant stars \citep{Politano_2010,Ivanova_2013}. 
Moreover, brown dwarfs do not form frequently enough to explain the occurrence rate of 
lithium-rich giants \citep{Cumming_2008}. Although intermediate mass 
($3.5\,M_\odot$ to $5\,M_\odot$) asymptotic giant branch stars can produce lithium
internally and transfer mass to a companion, they are also too rare to explain the number
of low-mass lithium-rich giant stars \citep{Karakas_2016}.
Nearby novae have been tentatively proposed as an explanation for lithium-rich giants \citep{Gratton_1989},  however the  novae rate is about two order of magnitudes different than the rate we infer for lithium-rich giant 
stars: \citet{Shafter_2017} find a novae rate of $50\,{\rm yr}^{-1}$, of which $\sim1/4$ are recurrent
novae and $\sim3/4$ have red giant donors \citep{Schaefer_2014}.
Although it is possible that only a fraction of novae could produce lithium-rich giants, 
no lithium-rich giants are known to show other abundance signatures that would be expected from classical novae \citep{Melo_2005}.
In summary, the formation rate we find excludes most 
external mechanisms proposed to explain the origin of lithium-rich giants.

Only a few proposed mechanisms remain, which are predominately associated with
stages of stellar evolution. However, our timescale analysis reveals that 
lithium enrichment is not predominately associated with either the luminosity
bump or the tip of the red giant branch (scenarios a and b), the only two
significant stellar evolution phases in a red giant star's evolution.
No lithium depletion timescale in either scenario is able to adequately account
for the observed distribution in stellar parameters (e.g., $\logg$), or reproduce
the observed fraction of core-helium burning lithium-rich giant stars.
This would suggest that lithium enrichment is not a consequence of single star evolution.

We find that the data are only consistent with scenario (c), where giant stars can become
lithium-rich at the core-helium burning phase or at a random time on the giant
branch. While we do find a relative weighting ($w_\mathrm{RGB}/w_\mathrm{CHeB} = 0.6$) and lithium
depletion timescale ($2\times10^6\,{\rm yr}^{-1}$) that can reproduce the
observations, we have yet to argue for any lithium enrichment mechanisms that 
could occur randomly on the giant branch, or at the start of the core-helium
burning phase.

\begin{figure*}
	\includegraphics[width=\textwidth]{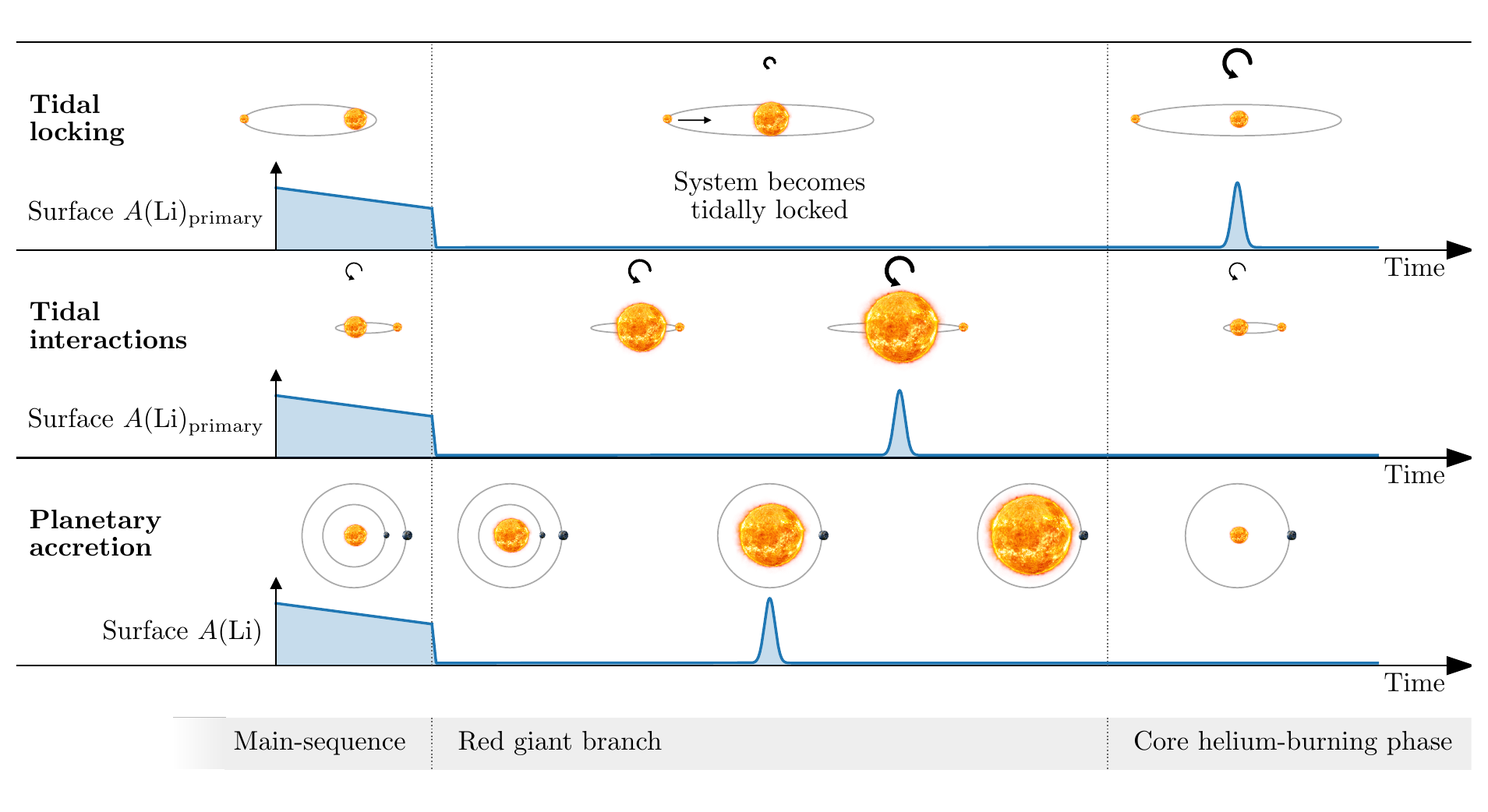}
	\caption{
		A schematic illustrating the mechanisms that can produce lithium-rich giants at the clump and at a random time on the giant branch. 
		Only tidal locking can produce lithium-rich giants in the core-helium burning phase. Tidal interactions can spin-up the primary when it is on the giant branch (spin is indicated by rotation arrows). Planetary accretion can also cause lithium enrichment (shown in blue) on the giant branch, where the time will depend on the planet's orbital radius. Schematic is not to scale in length or time.}
	\label{fig:diagram}
\end{figure*}

We argue that the mechanisms most consistent with scenario (c) are the 
accretion of a planet \citep{Siess_1999}, and the tidal spin-up from a binary companion \citep{Fekel_1993}. The accretion of a planet provides a reservoir
of unburnt lithium and acts as a mechanism to drive extra mixing that 
enables internal lithium production. 
A uniformly random time for lithium enrichment along the 
giant branch suggests
an event that occurs at a time that depends on the properties of that system.
Given a suitable distribution of exoplanet masses and periods, planet engulfment
is a plausible mechanism that could approximate a uniformly random lithium enrichment time 
on the giant branch.
However, planet accretion can only explain lithium-rich giants that do not have helium-burning cores. 
As a star evolves up the giant branch it expands in size until it reaches its maximum stellar radius at 
the tip of the giant branch, before contracting in radius over the next 
about $10^6~{\rm yr}$ as the star becomes a stable core-helium burning star. 
Any reasonably close-in planet (within $0.6\,{\rm AU}$ for a $1.3\,M_\odot$ solar-metallicity star) would have been accreted early on the giant branch (Figure~\ref{fig:diagram}).
Without 
introducing significant tidal decay to bring long-period planets close to the host star, planet accretion cannot explain lithium-rich giants with helium-burning cores.
In summary, planet accretion can only be responsible for up to about 20\% of lithium-rich giant stars.

The primary mechanism we propose for scenario (c) is tidal interactions between binary stars,
which can provide a consistent explanation for all
lithium-rich giant stars, including those with and without helium burning cores. 
Specifically, here we argue that internal lithium production at the start of the core-helium burning phase 
is an expected consequence of tidal locking in a binary system,
and tidal interactions could spin-up a red giant branch star at a near uniformly random time, depending on the properties of the binary system (Figure~\ref{fig:diagram}).
This mechanism is reliant on the internal production of lithium 
through the Cameron-Fowler mechanism \citep{Cameron_1971}, 
which is mixed to the surface of the giant star. In single star evolution, no net lithium is created without the introduction of extra
mixing.  Thermohaline mixing, driven by the burning of helium-3 outside the main hydrogen burning shell, can drive sufficient extra mixing after the luminosity bump to replenish some of the lithium lost during first dredge-up. This helium-3 captures an $\alpha$-particle to produce 
beryllium-7, which captures an electron to produce lithium-7. Without rapid mixing to move 
freshly  produced lithium-7 to a cooler region, lithium-7 easily captures a proton to 
form unstable beryllium-8, which then undergoes fission to helium-4. With rapid mixing 
the overall lithium abundance can increase inside a star: beryllium-7 is moved to a 
cooler region where electron-capture to lithium-7 can occur, but proton capture on 
lithium-7 does not occur. However, thermohaline mixing is insufficient to enhance lithium above the initial lithium abundance \citep{Lattanzio_2014}.
We can surmise that at least two conditions are required for lithium production inside red giants: there must be helium-3 available for the $^3{\rm He}(\alpha,\gamma)^7{\rm Be}$ reaction to occur, and the level of mixing inside a giant must be sufficient for beryllium-7 to be transported to cooler regions so that the $^7{\rm Be}(\beta^{-},\nu)^7{\rm Li}$ reaction can take place. 

We propose that differential rotation, enhanced by a binary companion, can induce sufficiently fast mixing to drive significant internal lithium production \citep{Costa_2002}. 
Let us first consider the case of a single star, without a binary companion.
Differential rotation is greatest when a star contracts to a core-helium burning star and the radius decreases by about a factor of ten or twenty \citep{Despain_1981}. We assume the contraction is homologous such that the moment of inertia  $I \propto MR^2$ and if we assume that the total angular momentum $J = I\Omega$ is conserved then $J \propto \Omega{}R^2$. Thus the spin $\Omega$ increases by a factor of 100 to 400 when core-helium burning begins. Rotation in centrally condensed stars \citep{Eddington_1929} generally causes perturbations of order $\Omega^2$ so that any rotationally driven mixing can be expected to increase with spin proportional to $\Omega^2$. 
Mixing due to rotation can be approximated as a diffusion process with a diffusion coefficient
\begin{equation}
D_{\rm mix} \approx \frac{L_{r}r^2}{\mathcal{M}_{r}g^{2}}\frac{\nabla_{ad}}{\nabla_{ad} - \nabla_{rad}}\Omega^2{f_i}
\end{equation}
\noindent{}which rises expectedly with $\Omega^2$, and it can be shown that $D_{\rm mix} > 10^{11}\,{\rm cm}^2\,{\rm s}^{-1}$ is necessary to enhance lithium significantly above its initial abundance \citep{Denissenkov_2004}. However, diffusion coefficients of $10^8\,{\rm cm}^2\,{\rm s}^{-1}$ to $10^9\,{\rm cm}^2\,{\rm s}^{-1}$ are found for normal red giant stars or core-helium burning stars of similar radii \citep{Denissenkov_2004,Palacios_2006}, 2 to 3 orders of magnitude below the ${D_{\rm mix} \approx 10^{11}\,{\rm cm}^2\,{\rm s}^{-1}}$ level required for lithium production. 
In other words, conservation of angular momentum ensures that the rotation of a typical single core-helium burning star is insufficient to enhance lithium.

A binary companion, however, can provide the additional angular momentum and subsequent higher
diffusion coefficient required for lithium production. We used a binary population 
synthesis code \citep{Hurley_2002} to model the tidal interactions and subsequent spin-up in binary systems.
For a representative case of a $1.5\,M_\odot$ primary star with a $1\,M_\odot$ companion,
the $1.5\,M_\odot$ primary expands to about $100\,R_\odot$ on its first 
ascent of the red giant branch. We require the two stars in the binary system 
to remain detached and so there
is a lower limit to the orbital period such that unstable Roche lobe overflow is avoided.

With a $1\,M_\odot$ companion the primary fills its Roche lobe in an orbit with a 
semi-major axis $a = 241\,R_\odot$ which corresponds to an orbital period of 
$P_{\rm orb} = 279\,{\rm days}$. In this binary system configuration the giant
primary is tidally locked early in its ascent of the giant branch so that at the point
of helium ignition the primary has spin
$P_{\rm spin} = P_{\rm orb} = 279~{\rm days}$. After the helium core flash the
primary shrinks to a radius of about $10\,R_\odot$. If we assume the collapse
is homologous then the angular velocity increases as $1/R^2$ or by a factor
of about 100. This spins up the primary to 
$P_{\rm spin} = 2.14\times10^5\,{\rm s}$ and an equatorial velocity of $182\,\rm{km\,s}^{-1}$.
Because $P_{\rm orb} = 279\,{\rm days}$ is the shortest period this representative binary system configuration
can accommodate while avoiding mass transfer, $182\,\rm{km\,s}^{-1}$ corresponds 
to the maximum spin-up of the primary during the core helium-burning phase.

As an upper limit to the binary period we require that the tidal synchronisation must be sufficient
before the primary reaches the tip of the giant branch. We find this requires initial
periods about ten times longer than the previously determined minimum orbital period ($P_{\rm orb} = 279\,{\rm days}$) or $P_{\rm init} = 7.64\,{\rm yr}$. 
At larger periods the tides are too weak to have an effect on the spin of the
primary.
After the helium core flash the primary shrinks and the core helium-burning star
in such a system increases its spin as usual, so that it is rotating  ten times slower than found for our shortest period configuration. This produces an equatorial
velocity of $18\,{\rm km\,s}^{-1}$, which is still fast compared to most giants. For a representative case of a $1.5\,M_\odot$ primary giant and a $1\,M_\odot$
companion, $18\,{\rm km\,s}^{-1}$ and $182\,\rm{km\,s}^{-1}$ represent the
lower and upper bounds of equatorial velocities expected from the tidal spin up of
a companion. This range of tidal spin-up would increase $D_{\rm mix}$ by up to a factor of about 6,500 ($D_{\rm mix} \approx 10^{12}$ to $10^{13}\,{\rm cm}^{2}{\rm s}^{-1}$), orders of magnitude above the requisite value to drive internal lithium production ($D_{\rm mix} \approx 10^{11}\,{\rm cm}^{2}{\rm s}^{-1}$).

This calculation demonstrates that tidal interactions between binary stars can drive 
lithium production in low mass red giant branch stars.
If we adopt a normal distribution in $\log{[P\,{\rm (days)}]}$ with a peak at $\log{P} = 5.03$ and $\sigma_{\log{P}} = 2.28$ \citep[as inferred from observations;][]{Raghavan_2010}, then up to 1 in 3 giants in a binary system could be affected by tidal spin up.
The effect of metallicity
is to change the maximum radius of the giant (to $71/87/155/175\,R_\odot$ for $Z = 0.0001/0.001/0.02/0.03$, respectively) 
while the effect on the 
radius of the core helium-burning star is to change by only about 10\% over the same
range. The range of periods over which stars become tidally synchronised on the red giant
branch varies accordingly as $P \propto R^{3/2}$ and the spin up is weaker by $R^2$, so the
final spin of a tidally synchronised core-helium burning star scales as $R^{-1/2}$.

Tidal interactions in binary systems would be consistent with an existing link proposed
between lithium enrichment and projected surface rotation. Giant 
stars are generally considered fast rotators if their projected surface rotation, 
$v\sin{i}$, exceeds $20\,{\rm km\,s}^{-1}$. The spectral resolution of \lamost\
prohibits us from detecting projected surface rotation below $120\,{\rm km\,s}^{-1}$.  
However, 103 of our lithium-rich giant stars appear in a study of stellar
rotation in \lamost\ \citep{Frasca_2016}. Of these 3 of 103 lithium-rich giants have
projected surface rotation significantly above the \lamost\ detection limit: between
$150$ and $260\,{\rm km\,s}^{-1}$. 
The remaining 100 lithium-rich giant stars have upper limits of less than $120\,{\rm km~s}^{-1}$.
We find that 140
of our lithium-rich giants were observed as part of the \apogee\ survey \citep{Abolfathi_2018}.
Only 5 of those 140 have measurements of $v\sin{i}$, ranging from $16\,{\rm km\,s}^{-1}$ to
$76\,{\rm km\,s}^{-1}$.  We also find 13 stars in common between our sample and
the fifth data release of the \rave\ survey \citep{Kunder_2017} and 11 of these 
have $v\sin{i}$ measurements ranging from $20\,{\rm km\,s}^{-1}$ and $41\,{\rm km\,s}^{-1}$.

We assume that selection effects in \rave, \lamost, and \apogee\ 
have no dependence on $v\sin{i}$, and that there are no systematic biases in $v\sin{i}$ between these surveys. We further assume that there are no other phenomena that would contribute to whether or not $v\sin{i}$ can be measured and infer that, if $v\sin{i}$ is not reported then $v\sin{i}$ is  so low that it could not be measured. This detection floor is $120\,{\rm km\,s}^{-1}$ for \lamost\ \citep{Frasca_2016}, about $10\,{\rm km\,s}^{-1}$ for \rave\ \citep{Siebert_2011}, 
and about $4\,{\rm km\,s}^{-1}$ for \apogee\ \citep{Deshpande_2013}.  With these assumptions we can na\"ively state that 1.2\% (3/256) of lithium-rich giants have $v\sin{i}$ that exceeds $120\,{\rm km\,s}^{-1}$. 
For the lithium-rich giants in \lamost\ that were also observed by \rave\ and \apogee,
we conclude that about 10.5\,\% (16/153) of lithium-rich giants are fast rotators ($v\sin{i} \gtrsim 20\,{\rm km\,s}^{-1}$). If the lithium depletion timescale is about $10^6\,{\rm years}$ and the spin-down timescale is about $10^5\,{\rm years}$ \citep{Tout_1992} then we can expect about 10\% of lithium-rich giants to show some level of enhanced rotation. We assume that the rotational velocities of stars experiencing spin-down are uniformly distributed between spun-up and  a representative level of $5\,{\rm km\,s}^{-1}$ for giants that have not experienced tidal interactions. 
For spun-up  giants we take $v\sin{i} \approx 18\,{\rm km\,s}^{-1}$ for stars in the widest tidally locked systems and $182\,{\rm km\,s}^{-1}$ for the closest. 
Given these assumptions the fraction of lithium-rich giants that we expect to have rotation above $120\,{\rm km\,s}^{-1}$  among the entire population (whether they were observed or not), is $f_{>150} = 1 - (150 - 5)/(182 - 5) = 0.18$.

The fraction of lithium-rich giants that we expect to have rotation exceeding $150\,{\rm km\,s}^{-1}$ is given by $f_{{\rm observe}\,>150} = \frac{t_{\rm spindown}}{t_{\rm depletion}} f_{\rm CHeB} f_{>150}$. Our analysis indicates that about 80\% of lithium-rich giants are core-helium burning giants ($f_{\rm CHeB} = 0.80$). Given our assumptions we conclude that about 1.5\% of lithium-rich giants with helium-burning cores should have surface rotation exceeding $150\,{\rm km\,s}^{-1}$. This is in good agreement with the 1.2\% (3/256) we observe from the na\"ive combination of \lamost, \rave, and \apogee\ data. If we only consider lithium-rich giants with $v\sin{i}$ measurements from \rave\ or \apogee\ then we can repeat this calculation for lower detectable rotational velocities. Taking $f_{>20} = 1 - (20 - 5)/(182 - 5) = 0.92$, we find 7.4\% of lithium-rich giants are expected to have surface rotation exceeding $20\,{\rm km\,s}^{-1}$. This, too, is in reasonable agreement with the 10.6\% (16/153) of lithium-rich giants observed to have rotation exceeding $20\,{\rm km\,s}^{-1}$ in the \rave\ and \apogee\ cross-matches. Given our assumptions,  our na\"ive treatment of the combination of multiple catalogues, and our ignorance on the inclination angle, we conclude that the level of projected surface rotation among lithium-rich giant stars is consistent with tidal spin-up by a binary companion.

We have argued that lithium production driven by tidal interactions is consistent
with the observations, but what evidence is there for binarity among lithium-rich
giants? Lithium-rich giants are not usually subject to repeat spectroscopic observations,
as a single high-quality spectrum is typically sufficient to derive detailed chemical
abundances and isotopic ratios, and most literature discussion to date has focussed 
on alternative hypotheses for lithium enrichment. For these reason, almost no 
lithium-rich giants have been repeatedly observed for radial velocity variations that would indicate binarity. This is largely the case for our sample: most sources have a single
epoch in \lamost\ and \rave, or just a few in \apogee. 
However, 8 lithium-rich giants were serendipitously discovered by an
exoplanet host star survey \citep{Adamow_2012,Adamow_2014,Adamow_2015,Adamow_2018,Deka-Szymankiewicz_2018,Zielinski_2012,Niedzielski_2016}, where multiple epochs of radial velocity measurements are available. Of those lithium-rich giants, 5/8 showed radial velocity
variations from just a few epochs: a curiously high fraction given the long orbital periods
where tides could drive lithium production.
A precise radial velocity study of a large number of lithium-rich giants 
is well-motivated, but the long orbital periods make such an endeavour expensive.

While tidal interactions between binary stars would provide a consistent explanation
for the data, some alternative explanations merit discussion. If planet accretion is
responsible for the 20\% of lithium-rich giants that are on their first ascent on the
giant branch, then the remaining 80\% could be explained by some mechanism associated
with the core helium flash that occurs at the tip of the red giant branch. The core
helium flash is a turbulent event, and extremely challenging to model accurately in
stellar evolution. Velocity fluctuations are necessarily suppressed, which limits
the inferences one can make on internal mixing and subsequent lithium production.
Let us consider that the data are explainable by scenario (c) where the 
physical mechanisms are planet engulfment and internal lithium production arising
from the core helium flash. If so, why do some stars become lithium-rich during
core helium flash, and some do not? Do tidal interactions have no impact on lithium
production, even though mixing is a key ingredient?
The answer to these questions is, quite obviously, that we do not
know because we cannot accurately model the helium flash. Even without a detailed
understanding of the core helium flash, we know that if the core helium flash were
to drive lithium production then a reservoir of helium-3 is necessary just outside the main hydrogen burning shell. However without differential rotation along the giant branch, most helium-3 will be depleted by extra mixing by the time the star reaches the tip of the giant branch (although a small amount will be continually produced by the hydrogen shell).
Presumably the helium-3 reservoir would be provided during the nucleosynthesis and mixing 
that results from the core helium flash. 

Although it may be challenging to unambiguously show that the core helium flash is
responsible for internal lithium production, our hypothesis does provide a number of
falsifiable predictions. 
If planet accretion is responsible for enhanced lithium in red giants without 
helium-burning cores, then an increase in beryllium is also expected \citep{Siess_1999,
Melo_2005}. If tidal interactions are responsible then we would expect the 
long-lasting extra mixing from the spin-up of a binary companion to fully deplete
the beryllium in a star \citep{Sackmann_1999}.
Similarly, for planet accretion to explain core helium-burning lithium-rich giants,
planet engulfment would have to occur when the star is at the tip of the red giant
branch, and the star has since evolved into the stable core-helium burning phase and
lithium will deplete within the next $\approx10^{6}\,{\rm yr}$. If tidal interactions -- and more specifically, tidal locking -- is the mechanism for lithium enrichment among core-helium burning stars, then we predict most core-helium burning lithium rich giants to have a binary companion.

\section{Conclusions} \label{sec:conclusions}

We report the discovery of \SampleSize\ lithium-rich giant stars identified
from low-resolution \lamost\ spectra, a sample size some 100 times larger than any other to date.
We find that lithium-rich giant stars occur more frequently with higher stellar 
metallicity, a result that reconciles tension between precise estimates of occurrence
rates in metal-poor environments, and significantly higher occurrence rate estimates
derived from the (metal-rich) field. We find that $80^{+7}_{-6}$\% of lithium-rich 
giant stars have helium-burning cores.

We find that lithium-rich giant stars cannot be solely explained by lithium production
at the luminosity bump, or at the tip of the red giant branch, suggesting that 
lithium-rich giants are not a consequence of single star evolution. However, we find
that the data are explainable by a scenario where stars either can become lithium-rich 
at a random time on the giant branch, or at the start of the core helium-burning
phase, and remain lithium-rich for about $2\times10^{6}\,{\rm yr}$. 
Given this lithium depletion timescale and an occurrence rate of lithium-rich giants, 
we estimate a formation rate of lithium-rich giants of $0.3\,{\rm yr}^{-1}$. This
formation rate rules out most proposed explanations as the dominant mechanism for lithium enrichment, including
stellar mergers, the engulfment of a brown dwarf, mass transfer from an asymptotic 
giant branch companion, and classical novae.

We argue that a combination of tidal interactions, and possibly planetary engulfment, 
are the most plausible mechanisms that are consistent with the data. However,
because a giant star increases in radius as it ascends the giant branch, 
planetary engulfment can only explain up to about 20\% of lithium-rich giants 
(e.g., those without helium-burning cores), as any core helium-burning lithium-rich
giant stars will have radii some 10-20 times lower than its radius on the giant branch, and therefore be unable to ingest giant planets without introducing significant tidal decay to bring long period planets close in.
We conclude that tidal interactions seem to be the most dominant and plausible
remaining explanation for lithium-rich giant stars.

We have shown that tidal interactions in binary systems can be strong enough
to drive internal mixing high enough such that lithium can be produced through the 
Cameron-Fowler mechanism. This effect is largest in a binary system where a giant star 
contracts in radius at the start of the core helium burning phase, consistent with
our scenario.
Those same conservation of angular momentum constraints demonstrate that the
requisite level of mixing cannot be achieved by a single star without a binary
companion. Although
there are observational biases, the expected projected rotational velocities 
resulting from tidal interactions are consistent with observations.

A prediction of our hypothesis is that nearly every lithium-rich giant star
with a helium-burning core has a binary companion, as most of these objects cannot be
explained by planet accretion. 
Similarly, unless 
the frequency of lithium-rich giants can be explained by an increased planet 
occurrence rate at higher metallicities,
then our hypothesis implies either an increasing binary fraction with increasing stellar 
metallicity, or that metal-rich stars are more affected by rotation induced by
tidal interactions.
Distinguishing between tidal spin-up and planet accretion may be possible for 
individual systems through high-resolution spectroscopic observations
to precisely measure beryllium and radial velocity variations. Planetary 
engulfment is expected to increase both beryllium and lithium \citep{Siess_1999,
Melo_2005}, whereas long-lasting extra mixing from the spin-up of a binary 
companion is expected to fully deplete the beryllium inside a star \citep{Sackmann_1999}.\\

\acknowledgments

We thank David W. Hogg (NYU).
A.~R.~C. is supported through an Australian Research Council Discovery Project under grant DP160100637. 
A.~Y.~Q.~H. is grateful to the community at the MPIA for their support and hospitality during the period in which much of this work was performed. 
A.~Y.~Q.~H. was supported by a Fulbright grant through the German-American Fulbright Commission and a National Science Foundation Graduate Research Fellowship under Grant No. DGE-1144469. 
M.~K.~N. and H.-W.~R. have received funding for this research from the European Research Council under the European Union's Seventh Framework Programme (FP 7) ERC Grant Agreement n. [321035].
C.~A.~T. thanks Churchill College for his fellowship and Monash University for hosting him as a Kevin Westfold distinguished visitor.
This work was supported by the GROWTH project funded by the National Science Foundation under PIRE Grant No 1545949.
The research leading to the presented results has received funding from the European Research Council under the European Community's Seventh Framework Programme (FP7/2007-2013)/ERC grant agreement (No 338251, StellarAges).
This research has made use of NASA's Astrophysics Data System.
This work has made use of data from the European Space Agency (ESA) mission Gaia (http://www.cosmos.esa.int/gaia), processed by the Gaia Data Processing and Analysis Consortium (DPAC, http://www.cosmos.esa.int/web/gaia/dpac/consortium). Funding for the DPAC has been provided by national institutions, in particular the institutions participating in the Gaia Multilateral Agreement.
Guoshoujing Telescope (the Large Sky Area Multi-Object Fiber Spectroscopic Telescope LAMOST) is a National Major Scientific Project built by the Chinese Academy of Sciences. Funding for the project has been provided by the National Development and Reform Commission. LAMOST is operated and managed by the National Astronomical Observatories, Chinese Academy of Sciences.
 This paper includes data collected by the Kepler mission. Funding for the Kepler mission is provided by the NASA Science Mission directorate.
 This paper includes data collected by the K2 mission. Funding for the K2 mission is provided by the NASA Science Mission directorate.

\vspace{5mm}
\facilities{LAMOST, Kepler, Gaia}

\software{
     \project{AstroPy}\ \citep{Astropy_2013,Astropy_2018},
     \project{numpy}\ \citep{Van-Der-Walt_2011},
     \project{scipy}\ \citep{Jones_2001},
     \project{matplotlib}\ \citep{Hunter_2007},
     \project{The Cannon}\ \citep{Ness_2015,Casey_2016b,Ho_2017a,Ho_2017b}
}    
     
\bibliographystyle{aasjournal}
\bibliography{eneloop}


\end{document}